\documentclass[12pt]{article}
\usepackage{a4wide}
\usepackage{graphicx,rotating}
\usepackage{float}
\usepackage{color}\usepackage{wrapfig}
\usepackage{subfigure}\usepackage[latin1]{inputenc}
\begin{document}
\pagestyle{empty}
\begin{center}\Large {\bf  Turbulence characteristics of
    electron cyclotron and ohmic heated discharges}~\\*[0.5cm]
  \normalsize I. Pusztai$^{(1)}$, S. Moradi$^{(1)}$,
  T. F\"ul\"op$^{(1)}$, N. Timchenko$^{(2)}$ \\ {\it\small $^{(1)}$
    Department of Applied Physics, Nuclear Engineering, Chalmers
    University of Technology and Euratom-VR Association, G\"oteborg,
    Sweden\\ $^{(2)}$ Institute of Tokamak Physics, NRC ``Kurchatov
    Institute'', 123182, Kurchatov Sq. 1, Moscow, Russia
  }\\\end{center}
\begin{abstract}
  Turbulence characteristics of electron cyclotron (EC) and ohmic
  heated (OH) discharges has been analyzed by electrostatic gyrokinetic
  simulations with {\sc gyro} [J. Candy, R.E. Waltz, Journal of
  Computational Physics {\bf 186}, 545-581 (2003)] aiming to find
  insights into the effect of auxiliary heating on the
  transport. Trapped electron modes are found to be unstable in both
  OH and the EC heated scenarios. In the OH case the main drive is
  from the density gradient and in the EC case from the electron
  temperature gradient. The growth rates and particle fluxes exhibit
  qualitatively different scaling with the electron-to-ion temperature
  ratios in the two cases. This is mainly due to the fact that the
  dominant drives and the collisionalities are different.  The inward
  flow velocity of impurities and the impurity diffusion coefficient
  decreases when applying EC heating, which leads to lower impurity
  peaking, consistently with experimental observations.
\end{abstract}
\section{Introduction}
Even if there is a wealth of experimental data and theoretical models
relating to the effect of auxiliary heating on transport there are
still many open issues regarding the sign and magnitude of the
transport and its parametric dependencies.  One example of this is the
experimental observation that the density profiles of electrons and
impurities depends on the auxiliary heating. Results from many
different devices have shown a flattening effect of electron cyclotron
resonance heating (ECRH) on the electron density
\cite{d3d,jt60,tcv}. Furthermore, impurity accumulation also can be
reduced by central ECRH \cite{asdex,tcv1,warsaw}. However, in some
parameter regions ECRH does not affect the electron or impurity
density profiles, or even peaking of these profiles is observed
\cite{peakangioni1,peakangioni2}. The physical mechanism giving rise
to the differences is not clearly identified, although it seems that
collisionality plays a crucial role in determining the particle
transport \cite{angionidensitypeak,peaking1,peaking2,peaking3}, while
the electron-to-ion temperature ratio \cite{peakangioni2}, and the
density and temperature scale lengths \cite{peakangioni1} are also
important. In order to be able to make predictions confidently for
future fusion devices understanding of the underlying transport
processes is necessary for a wide range of these parameters.  In
particular, to understand how and why the transport processes are
different in the ohmic and EC heated discharges it is important to
analyze the turbulence characteristics and scalings with key
parameters such as collisionality, electron-to-ion temperature ratio,
and density and temperature scale lengths.

The aim of the paper is to calculate the turbulence characteristics
and the corresponding particle and energy fluxes for two similar
experimental scenarios, one with ECRH and one with only ohmic
heating. The steady-state impurity density gradient for trace
impurities will also be calculated.  As there is a consensus that the
transport in tokamak core plasmas is mainly dominated by transport
driven by drift wave instabilities, we focus on these instabilities
and use quasilinear numerical simulations with the {\sc gyro} code
\cite{gyro} to calculate the turbulence characteristics. As shown in
\cite{dannertjenko,casati} the quasilinear electrostatic approximation
retains much of the relevant physics and reproduce the results of
nonlinear gyrokinetic simulations for a wide range of parameters.

The experimental scenarios from T-10 are well-suited for the study we
perform. One of the advantages is that there are measurements of
turbulence characteristics on T-10 which can be used to compare with
our theoretical calculations.  The selected discharges have
hot-electron plasma, relatively high density and collisionality;
analyzing experimental scenarios from it gives insights into a range
of parameters that have not been studied before.

Our results show that the dominant instability is trapped electron
mode in both OH and the EC heated scenarios. As expected, the
collision frequency plays an important role stabilizing the trapped
electron mode driven turbulence in both cases. The EC heated case is
more strongly suppressed for lower collisionalities. This is due to the
drop of the electron temperature gradient drive which is stabilized by
collisions. The growth rates and particle fluxes exhibit qualitatively
different scalings with the electron-to-ion temperature ratios in the
two cases. This is mainly due to the different collisionalities, but
in the case of the electron particle flux also the difference in
density gradients contributes. Sensitivity scalings for electron
density and electron temperature gradients show that both of these
drives are present in the investigated experimental scenarios. The
inward flow velocity of impurities and the impurity diffusion
coefficient decreases when applying EC heating, which leads to lower
impurity peaking, consistently with experimental observations.

The remainder of the paper is organized as follows.  In
Sec.~\ref{sec:scenario}, the experimental scenario and theoretical
modeling are described.  In Sec.~\ref{sec:results}, the simulation
results are presented and interpreted. We present the instability
properties, the background ion and electron transport fluxes and
convective and diffusive transport of a trace impurity species.
Finally, the results are summarized in Sec.~\ref{sec:conclusions}.


\section{Scenario and theoretical modeling}\label{sec:scenario}
\subsection{Description of the T-10 discharges}
\begin{figure}[htbp]
\begin{center}
\includegraphics[width=0.9\linewidth]{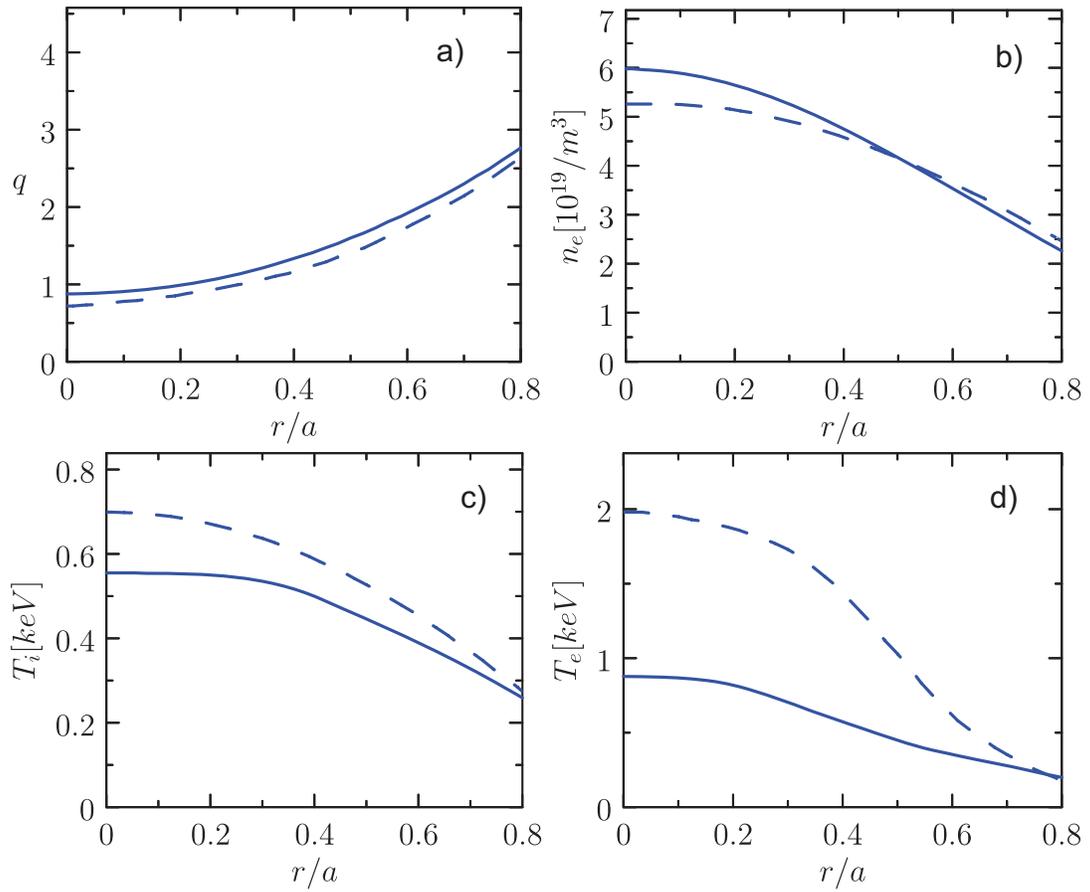}
\caption{Safety factor (a),  electron density (b) and ion and electron
  temperatures (c and d respectively) for the two cases (OH solid, EC dashed).}
\label{parameters1}
\end{center}
\end{figure}

The T-10 plasma has circular cross-section with major radius $R =
1.5\;\rm{m}$ and minor radius $a = 0.3\;\rm{m}$. The plasma current in
the discharges was $I_p=200\;\rm{kA}$ and the toroidal magnetic field
was $B_T=2.4\;\rm{T}$.  The effective charge in the comparatively high
density discharges we studied was rather low $Z_{\rm eff}=1.2$. We
study two typical experimental scenarios from T-10, one with
$1\;\rm{MW}$ electron cyclotron heating (EC), and one with only Ohmic
heating (OH).  The plasma parameter profiles used in our study are
given in Fig.~\ref{parameters1}. Carbon is an intrinsic impurity in
all the discharges and here it is taken to be $n_C/n_e =0.67\%$ unless
otherwise is stated.

In discharges with on-axis EC heating, the electron density was found
to decrease in the plasma center.  There is sawteeth activity in the
central part of the plasma in both discharges and the inversion radius
is around $7\,\rm{cm}$. Therefore we will concentrate our studies
outside $r/a=0.4$.

\subsection{Gyrokinetic modeling}
\label{sec:gkmod}

The linear gyrokinetic dynamics of all ion specii (deuterium, carbon
and a trace species with the concentration $n_Z/n_e=10^{-5}$) and
electrons has been simulated using the {\sc gyro} code. We consider only
electrostatic fluctuations relevant for low $\beta$ and circular
geometry. All the species are kinetic, and we include parallel
compressibility and electron-ion collisions. We note that the ion-ion
collisions were found to be unimportant even in the highest
collisionality regions we studied. The carbon impurity is treated
self-consistently. Unless otherwise specified, the following
conventions and units are employed throughout this paper. Frequencies
and growth rates are normalized to $c_{s}/a$ where $c_{s}
=\sqrt{T_e/m_i}$ is the ion sound speed, $a$ is the plasma minor
radius, and $i$ is the main ion species. The fluxes are normalized to
the flux surface average of $ k_\theta\rho_s|e\phi/T_e|^2$, where
$\phi$ is the amplitude of the fluctuating electrostatic potential,
$k_\theta$ is the poloidal wave number, $\rho_s=c_s/\omega_{ci}$ is
the ion sound Larmor radius with $\omega_{ci}=eB/m_i$ the ion
cyclotron frequency. The radial scale lengths are defined as
$L_{n\alpha}=-[\partial (\ln{n_\alpha})/\partial r]^{-1}$ and $L_{T\alpha}=-[\partial
(\ln{T_\alpha})/\partial r]^{-1}$, where $\alpha$ denotes the particle species.

The magneto-hydrodynamic equilibrium -- including the Shafranov shift
of the circular flux surfaces -- is calculated by the {\sc astra}
code~\cite{astra}. As the plasma rotation in the T-10 tokamak is weak
it is  neglected in our simulations.

The linear {\sc gyro} simulations were carried out using flux-tube
(periodic) boundary conditions, with a 128 point velocity space grid
(8 energies, 8 pitch angles and two signs of velocity), and the number
of poloidal grid points along particle orbits is 14 for passing
particles. The location of the highest energy grid point is at $m_i
v^2/(2T_i)=6$.  In the cases investigated here $\rho_\ast=\rho_s/a$
varied in the range of $1.0-9.4\cdot 10^{-3}$.

In the following chapter we present the linear frequencies and growth
rates from {\sc gyro} simulations to identify microinstabilities present in
the experimental cases. Then for mid-radius and a representative wave
number we perform parameter scalings to investigate the effect of
collisions, temperature ratio, electron density and ion temperature
gradients on frequencies and linear particle and energy
fluxes. Finally, the diffusion and particle flow and the zero flux
density gradient of a trace impurity species is studied through
impurity density gradient scalings.


\section{Instabilities and transport}\label{sec:results}
\subsection{Instability characteristics} 
\begin{figure}[H]
\begin{center}
\includegraphics[width=0.9\linewidth]{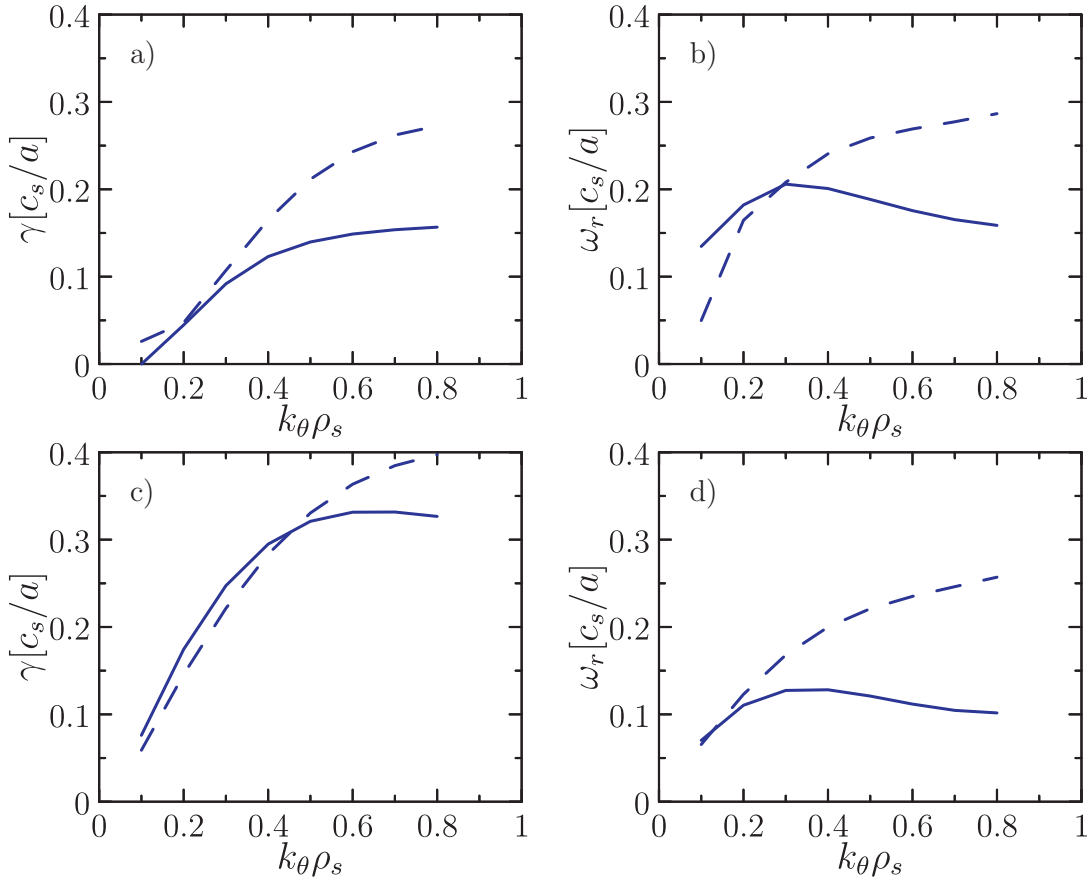}
\caption{Growth rates (a,c) and real frequencies (b,d) of the
  instability at $r/a=0.5$ (OH solid, EC dashed). Lower figures (c,d)
  are with collisionality switched off. }
\label{fig05}
\end{center}
\end{figure}

Figure \ref{fig05} shows the growth rates and real frequencies of the
instability as a function of $k_\theta\rho_i$ at $r/a=0.5$ both in the
collisionless case (a,b) and with collisions included (c,d).  The
linear simulations were performed for a range of
$k_\theta\rho_s$-values using the \emph{Maxwell dispersion matrix
  eigenvalue solver} method of {\sc gyro} to solve linear gyrokinetic (GK)
equations. This method is capable of finding all the unstable roots of
the Maxwell dispersion matrix, even the sub-dominant ones.

The real part of the mode frequency is positive in all cases that
suggests that the unstable modes are Trapped Electron (TE) modes. This
is confirmed by the fact that no unstable mode could be found if the
non-adiabatic electron response was switched off. It is known that the
non-adiabatic electron response can increase the ion temperature
gradient (ITG) mode growth rates, but the only unstable root is found
to be a TE mode in both the OH and EC cases.  This is in agreement
with the experimental observations of the turbulent characteristics in
high density cases similar to the ones studied here \cite{sofia}.

Trapped electron modes can be destabilized by both electron density
and electron temperature gradients \cite{angionitem}. The normalized logarithmic
density and temperature gradients at $r/a=0.5$ were $a/L_{ne}=1.47$
and $a/L_{Te}=2.62$ in the OH, and $a/L_{ne}=1.08$ and $a/L_{Te}=3.96$
in the EC case. One might expect that the higher growth rates in the
EC case (dashed line in Fig.~\ref{fig05}a) are due to the high value
of $a/L_{Te}$, but from Fig.~\ref{fig05}c it becomes clear that the
difference is mainly due to the effect of different collision
frequencies, as without collisions the growth rates are rather
similar.

\subsection{Effect of collisions}
\label{sec:collef}
\begin{figure}[H]
\begin{center}
\includegraphics[width=0.9\linewidth]{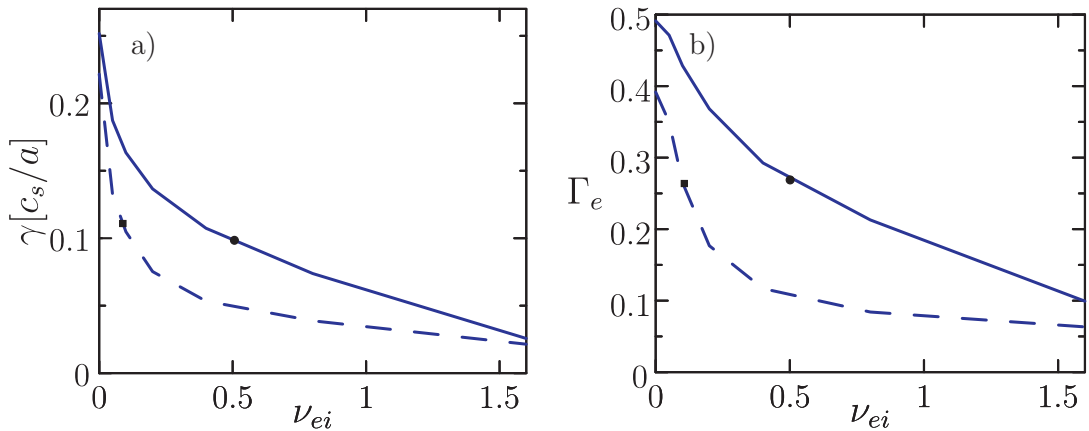}
\caption{Growth rate (a) and linear electron particle flux (b) of the
  instability obtained with linear {\sc gyro} calculations, as a function of
  electron-ion collision frequency. Solid curve: OH, dashed curve:
  EC. The markers correspond to the experimental value of the
  collision frequency ($r/a=0.5$, $k_\theta\rho_s=0.3$). }
\label{collscale}
\end{center}
\end{figure}

It is interesting to note that the growth rate of the instabilities
are reduced by the collisions as it can be seen on Fig.~\ref{fig05}
comparing the upper (with collisions) and lower (without collisions)
figures.  This is due to the collisional de-trapping of trapped
electrons and has been noted before in e.g. Ref.~\cite{chh} where the
growth rate of a Dissipative Trapped Electron (DTE) mode in the long
wavelength limit is found to be $\gamma\sim\epsilon^{3/2}\omega_{\ast
  e}^2\eta_e/\nu_{ei}$, where $\epsilon=r/R$ is the inverse aspect
ratio, $\omega_{\ast e}=k_\theta T_e/(eBL_{ne})$ is the electron
diamagnetic frequency, $\eta_e=L_{ne}/L_{Te}$, and $\nu_{ei}$ is the
electron-ion collision frequency. Our cases are quite similar to the
DTE region, the parameter $(\nu_{ei}/\epsilon)/ |\omega|$ is typically
much higher than one due to the modest temperatures and the high
aspect ratio. On the other hand, the mentioned expression for the DTE
growth rate in \cite{chh} is obtained excluding magnetic drifts
therefore we do not expect to find the same parametric dependence on
collision frequency. The sensitivity of the growth rate and electron
particle flux to the variation of the collision frequency in the OH
and EC cases is illustrated on Fig.~\ref{collscale} a and b
respectively. The curves of the figure, as in all figures henceforth,
correspond to mid-radius $r/a=0.5$ and $k_\theta\rho_s=0.3$ which in
these TE mode cases does not correspond to the fastest growing mode
being at higher wave numbers, but to a typical maximum of nonlinear
fluxes.

In previous trapped electron mode studies \cite{angionitem} it was
found that the electron temperature drive of the TE modes is strongly
suppressed as the collision frequency increases, while the density
gradient drive can remain for higher collisionalities. From this
perspective -- although our experimental cases are not extreme
examples for pure electron density or temperature driven TE modes --
the OH case is more similar to the density gradient driven TE mode
while the EC case having higher temperature gradient and lower density
gradient is mainly driven by $a/L_{Te}$. For lower collision
frequencies -- around the experimental value of $\nu_{ei}$ in the EC
case -- the EC growth rate strongly decreases with increasing
collisionality as the $a/L_{Te}$ drive is suppressed, but for higher
collisionalities the mode is not completely stabilized due to the
finite density gradient drive. The growth rate in the OH case or low
collisionalities does not exhibit so strong dependence on $\nu_{ei}$,
and that is what we expect in the density gradient driven TE case.

The collisional stabilization of the modes also affects the electron
particle fluxes (Fig.~\ref{collscale} b), which exhibit qualitatively
similar dependence on collision frequency as the growth rates for most
of the plotted collisionality region. 

Regarding both the growth rates and particle fluxes the experimental
values are approximately the same in the two experimental scenarios,
but this seems to be a coincidence considering the strong dependence
of these quantities on collision frequency (and accordingly even
stronger dependence on electron temperature).


\subsection{Temperature ratio effects}

\begin{figure}[htbp]
\begin{center}
\includegraphics[width=0.9\linewidth]{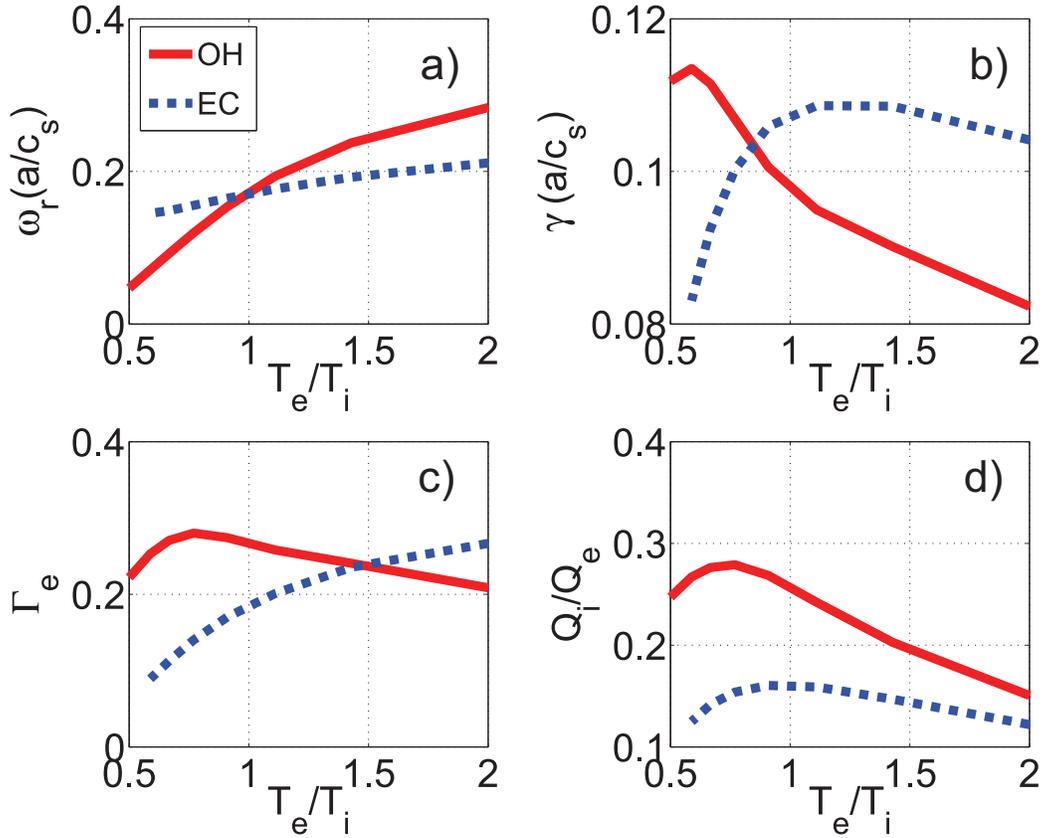}
\caption{$T_e/T_i$-scan of real frequencies (a) and growth rates (b) of
  the instabilities, the electron particle flux (c), and the ratio of ion
  and electron energy fluxes (d) for $r/a=0.5$ and $k_\theta
  \rho_s=0.3$.  The OH case is shown by solid lines, the EC with
  dashed lines. }
\label{figteti}
\end{center}
\end{figure}

It is reasonable to assume that one of the most important parameters
that causes the differences between the OH and EC plasmas is the
electron-to-ion temperature ratio. This parameter is indeed quite
different for these cases; at $r/a=0.5$, $T_e/T_i=1.01$ in the OH and
$T_e/T_i=1.95$ in the EC case. However, the effect of $T_e/T_i$ on the
properties of the instability and the transport is qualitatively
different in the two cases, as it can be seen in Fig.~\ref{figteti},
where the real frequencies and growth rates of the instability are
plotted together with the electron particle flux and $Q_i/Q_e$ as a
function of $T_e/T_i$. [We note that these scalings are performed
keeping the electron-ion collision frequency ($\nu_{ei}\propto
T_e^{-3/2}$) and the normalized ion sound Larmor radius
($\rho_s/a\propto T_e^{1/2}$) fixed.]  

In the OH case the growth rate of the instability strongly decreases
with this parameter almost on the whole temperature ratio region
plotted, while the dependence is much weaker in the EC case, in the
experimentally relevant regions. It is interesting to note that the
slope of the EC growth rate curve is positive for the experimental
value of the temperature ratio in the OH case.  The slope of the
$\Gamma_e(T_e/T_i)$ curve is negative in the OH case while it is
positive for the EC case. Thus -- although the experimental electron
particle flux values happen to be the same -- the flux in the two
cases exhibit qualitatively different behavior.  The shape of the
ion-to-electron energy flux ratio curves are similar in the two cases,
the EC case being lower due to the higher electron energy flux
corresponding to the stronger heating of electrons.

\begin{figure}[htbp]
\begin{center}
\includegraphics[width=0.9\linewidth]{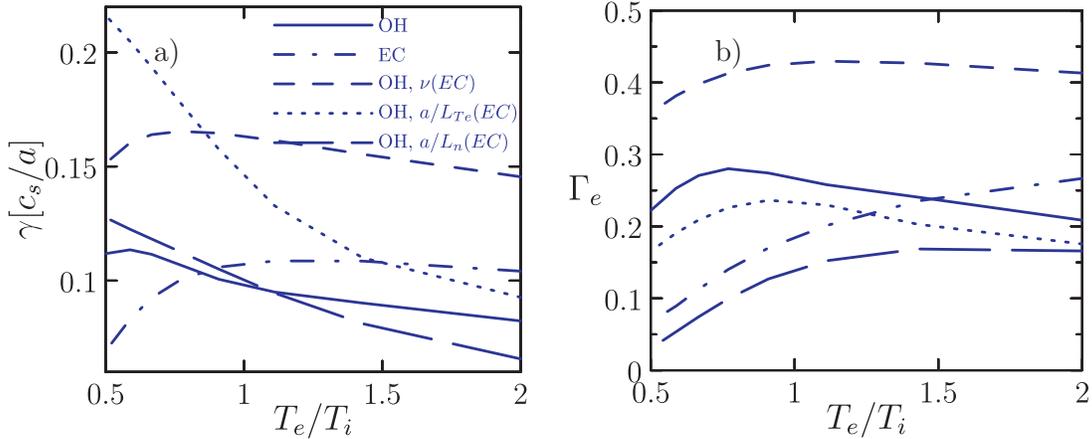}
\caption{Electron-to-ion temperature ratio scalings of growth rates (a)
  and electron particle fluxes (b) for the original experimental cases
  (solid: OH, dash-dotted: EC) and in cases where all the parameters
  are taken from the OH case, expect one, which is taken from the EC
  case. This parameter is chosen to be $\nu_{ei}$ (dashed), $a/L_{Te}$
  (dotted) and $a/L_n$ (long dashed). }
\label{figteti2}
\end{center}
\end{figure}

In order to determine which parameter causes the qualitatively
different $T_e/T_i$ scalings between the OH and EC scenarios we
performed simulations where all parameters were identical to those of
the OH case, except one, which we set to the corresponding value in
the EC case. We expect that the TE mode growth rates are mainly
affected by the density and electron temperature scale length and, as
we saw in the section \ref{sec:collef}, the collisionality. The result
of these simulations are shown in Fig.~\ref{figteti2}. The growth rate
in the OH case (solid line, figure a) is mainly decreasing with
$T_e/T_i$ and this trend is even emphasized when we changed $a/L_{Te}$
(dotted) or $a/L_{n}$ (long dashed) to their value in the EC
case. However, when the collision frequency was changed (dashed) the
behavior of the growth rate curve became somewhat similar to that in
the EC case (dash-dotted); the region for lower temperature ratios
where the OH case showed increase in this parameter widened and the
negative slope of the curve after the maximum growth rate is
reduced. This effect of the collision frequency can be due to that for
lower collisionalities -- as it is in the EC case -- the
temperature gradient drive of the TE modes is more pronounced.

One might expect that the difference in the particle fluxes have the
same origin as for the differences in the growth rates; that would
mean that the modified collisionality particle flux (dashed curve,
Fig.~\ref{figteti2} b) should exhibit similar behavior to the EC flux
(dash-dotted). This is partly true, as the positive slope region of
the modified collisionality flux become somewhat wider and for higher values of
$T_e/T_i$ the negative slope of the curve decreased. However, changing
the density gradient (long dashed curve) shifted the shape of the
particle flux curve closest to the EC heated case.  From this we can
conclude that the qualitative behavior of the temperature ratio
scaling of the growth rates is mainly affected by collisions, while
for the electron particle flux the electron density gradient is also
an important parameter from this aspect.

\subsection{Sensitivity to density and electron temperature gradients}
\begin{figure}[htbp]
\begin{center}
\includegraphics[width=0.9\linewidth]{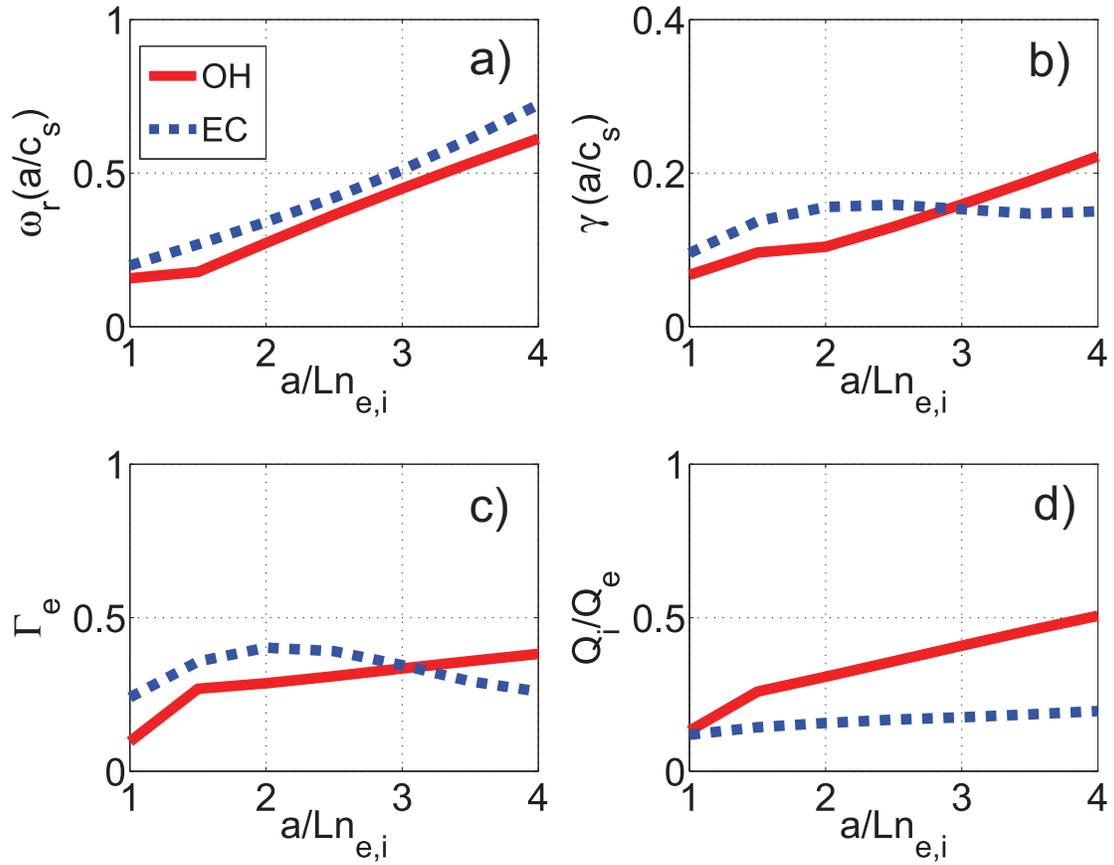}
\caption{$a/L_n$-scan of real frequencies (a) and growth rates (b) of
  the instabilities, the electron particle flux (c), and the ratio of ion
  and electron energy fluxes (d) for $r/a=0.5$ and $k_\theta
  \rho_s=0.3$.  The OH case is shown by solid lines, the EC with
  dashed lines. }
\label{figaln}
\end{center}
\end{figure}

Figure \ref{figaln} shows the real frequencies and growth rates of the
instability, together with the electron particle flux and $Q_i/Q_e$ as
a function of the logarithmic density scale length $a/L_{n}$.

Recalling that the logarithmic density gradients were $a/L_{ne}=1.47$
in the OH, and $a/L_{ne}=1.08$ in the EC case, for $r/a=0.5$, we find
that around these values the growth rate as a function of $a/L_n$
increases in both cases. This is not in contradiction with our
previous statement that the EC case is more similar to a temperature
gradient driven TE and the ohmic case is to a density gradient driven
TE since both of them has some contributions from both drives.  There
is no qualitative difference in the $\Gamma_e(a/L_n)$ curves in the
experimentally relevant region, but interestingly if we allow higher
density gradients we find that above $a/L_n\approx 2.2$ the electron
flux decreases with $a/L_n$ in the EC case, in contrast to the OH case
which always drives higher particle flux for higher density
gradient. This difference in the fluxes is related to the different
behavior of the growth rates for higher density gradients; in the EC
case $\gamma$ saturates, while it is steadily growing with $a/L_n$ in
the OH case. The sensitivity of the energy flux ratio to the density
gradients is more pronounced in the OH case than the EC case, but in
both cases $Q_i/Q_e(a/L_n)$ has a positive slope.

\begin{figure}[H]
\begin{center}
\includegraphics[width=0.9\linewidth]{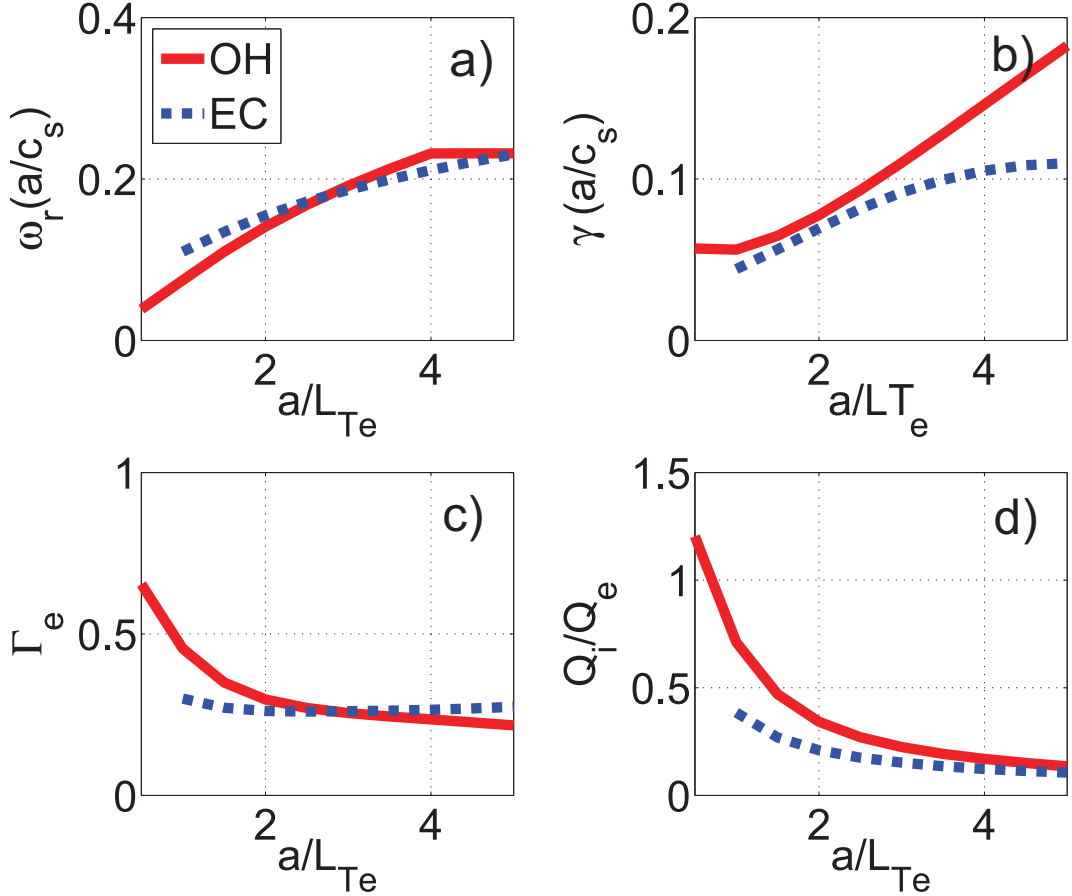}
\caption{$a/L_{Te}$-scan of real frequencies (a) and growth rates (b) of
  the instabilities, the electron particle flux (c), and the ratio of ion
  and electron energy fluxes (d) for $r/a=0.5$ and $k_\theta
  \rho_s=0.3$.  The OH case is shown by solid lines, the EC with
  dashed lines.}
\label{figalTe}
\end{center}
\end{figure}

Figure \ref{figalTe} shows the real frequencies and growth rates of the
instability, together with the electron particle flux and $Q_i/Q_e$ as
a function of the logarithmic electron temperature scale length $a/L_{Te}$.
Keeping in mind that $a/L_{Te}=2.62$ in the OH and $a/L_{Te}=3.96$ in
the EC case, we can see that the growth rate in the OH case are more
strongly affected by a change in the electron temperature gradients
than that in the EC case. This can be seem somewhat counter-intuitive
while we state that the EC case is mainly driven by the electron
temperature gradient. However, we should consider that in both of
these cases the TE mode is not exclusively driven by either of the
gradients, but both of them with different weights; clearly, increasing
the $a/L_{Te}$ in the OH case leads to that the case becomes more
strongly driven by electron temperature gradients.  Interestingly,
although in the OH case the growth rate increases rapidly when the
temperature gradient is increased and the $Q_i/Q_e$ ratio is
decreasing due to the higher electron energy flux, the electron
particle flux decreases with this parameter. In the EC case the
electron particle flux is approximately constant over the plotted
region, having a slight positive slope at the experimental value of
$a/L_{Te}$.

\subsection{Impurity peaking factor}

In similar experiments as the one studied here, a short argon gas-puff
was applied in a stationary phase of both OH and EC heated discharges,
and the evolution of the density of Ar$^{+16}$ impurity (with density
$n_z/n_e\approx 0.3\%$) was studied \cite{warsaw}. In discharges with
on-axis EC heating, the argon density was found to decrease in the
plasma center. The argon density reduction was proportional to the
total input power. In this work we have studied the transport of a
trace impurity with different charge in both EC and OH discharges. Our
results indicate that the peaking factor becomes lower in the EC case,
but it is still positive.

In experimental work, the particle diffusivity is often separated into
a diffusive part and a convective part  
$$
\Gamma_z=-D_z\frac{\partial n_z}{\partial r}+ V_z n_z
$$ where $D_z$ is the diffusion coefficient and $V_z$ is the
convective velocity. This separation of the flux into diffusion and
convection can be done in the trace limit of impurity concentrations,
because then the turbulence remains unaffected by the presence of the
impurity, and the impurity flux varies linearly with the impurity
density gradient. The particle flux of higher concentration minority
ions, bulk ions or electrons depends non-linearly on their density
gradient as it can affect the growth rates of the mode.

\begin{figure}[htbp]
\begin{center}
\includegraphics[width=0.9\linewidth]{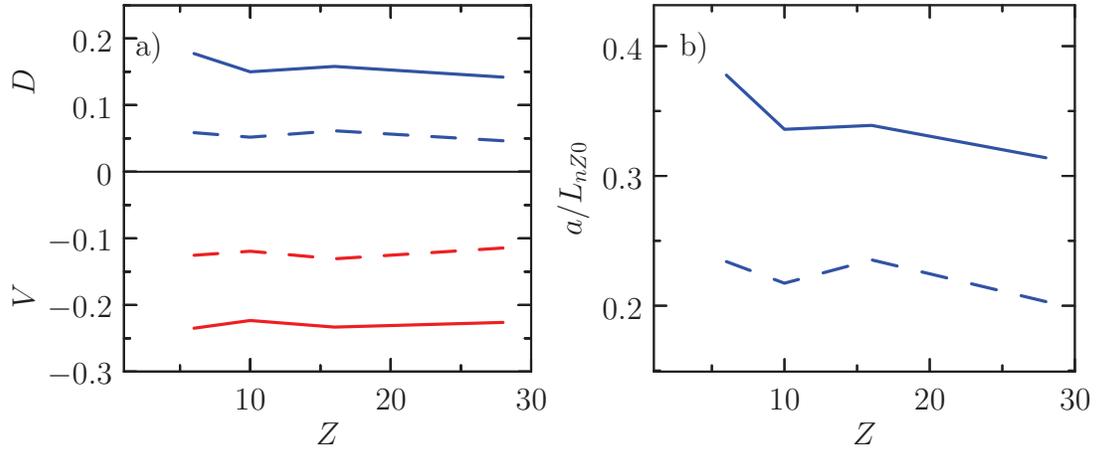}
  \caption{a) Impurity convective flux V (arb.~u., red curves) and
    diffusion coefficient D (arb.~u., blue curves) for different
    impurity charge numbers (OH solid, EC dashed). b) Impurity peaking
    factor.}
\label{figvd}
\end{center}
\end{figure}

The amplitude of the perturbed quantities cannot be calculated in
linear simulations, accordingly the convective velocity and
diffusivity are plotted in arbitrary units in
Fig.~\ref{figvd}~a, still allowing the comparison of the
two experimental cases. In Fig.~\ref{figvd} b the steady state density
gradient (or peaking factor) of impurities, $a/L_{nZ0}$  is shown. The peaking
factor is calculated from the criterion $\Gamma_z(a/L_{nZ0})=0$
(assuming the non-turbulent particle fluxes and the impurity sources
in the core negligible). All three quantities plotted in
Fig.~\ref{figvd} exhibit a rather weak dependence on impurity
charge. Furthermore for all three quantities, the ratio of their
values for the two different experimental cases is almost independent
of $Z$.

\begin{figure}[htbp]
\begin{center}
\includegraphics[width=0.9\linewidth]{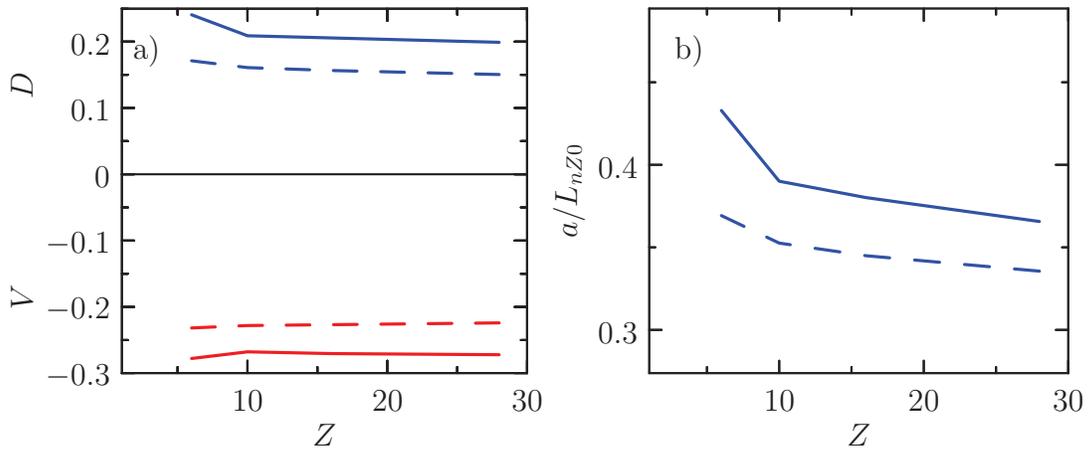}
  \caption{ Impurity transport \emph{without parallel ion
      compressibility}. a) Impurity convective flux V (arb.~u., red
    curves) and diffusion coefficient D (arb.~u., blue curves) for
    different impurity charge numbers (OH solid, EC dashed). b)
    Impurity peaking factor.}
\label{figvdnopar}
\end{center}
\end{figure}

The impurity peaking factor is approximately $2/3$ times lower in the
electron cyclotron heated case, consistently with the experimentally
observed decrease of impurity content with EC heating. The inward
pinch velocity is approximately $1/2$ times lower in the EC heated
case than in the OH case and the ratio of diffusion coefficients in
the EC and the OH case is lower being around $1/3$, which leads to the
lower peaking factor in the EC case.

The reduction of the peaking factor and in some cases even reversal of
the impurity flux from inward to outward in the presence of ECRH has
been noted before also in other experiments \cite{asdex,tcv1}. This
was partly explained by the fact that the fluctuation of the parallel
velocity of impurities along the field lines can generate an outward
radial convection for TE-modes \cite{ap}. We can identify the
contribution of this effect in our cases by switching of the parallel
ion dynamics in the gyrokinetic simulations, see the corresponding
plots of $V$, $D$ and impurity peaking factor in
Fig.~\ref{figvdnopar}. The difference between the two experimental
cases is smaller without parallel ion dynamics which means that,
indeed, the contribution of the parallel ion compressibility drives
the system away from peaked impurity profiles.
However, the value is positive for both cases, although in experiments
the impurities have a hollow profile corresponding to negative peaking
factor. The reasons for the discrepancy might be due to that the
effect of the Ware pinch is not considered, and we have only one wave
number in a linear simulation instead of a whole range of interacting
modes as in a nonlinear simulation. It should also be mentioned that
gyrokinetic simulations perform usually better in terms of energy
fluxes than for particle fluxes~\cite{angionifable,waltzcasati}, thus
we do not expect perfect agreement between the experimental and
simulated peaking factors.

 \section{Conclusions}\label{sec:conclusions}

 We compared the transport characteristics in electron cyclotron
 heated and purely ohmic plasmas on the T-10 tokamak using linear
 eigenvalue solver gyrokinetic simulations with the {\sc gyro} code. The aim
 was to obtain insights to the effect of electron cyclotron heating on
 the microinstabilities driving the turbulence, the corresponding
 particle and energy fluxes, and on the impurity particle transport.

 The only linearly unstable mode found in these experimental cases is
 a trapped electron mode. The frequency of collisional de-trapping is
 typically much higher than the mode frequency in these cases,
 accordingly the instabilities exhibit dissipative TE mode features;
 they are stabilized by collisions. However the modes are not
 completely stabilized by the collisions similarly to what was
 previously found in \cite{angionitem} for density driven trapped
 electron modes. The higher linear growth rates found in the EC case
 are mainly due to the lower collision frequency in this high electron
 temperature plasma, and it is not an effect of the higher electron
 temperature gradient.

 The dependence of electron particle flux on $T_e/T_i$ is
 qualitatively different in the two cases; in the OH case the electron
 particle flux decreases with this parameter, while it increases in
 the EC case. This behavior can be understood noting that the growth
 rate in the OH case decreases with increasing $T_e/T_i$, but in the
 EC case the dependence of the growth rate on this parameter is much
 weaker.  The TE mode growth rate in the EC case strongly increases
 with increasing density gradient, while the growth rate in the OH
 case is almost independent on this parameter for the experimentally
 relevant region $a/L_{ne}\sim 1-1.5$. In spite of the differences in
 the growth rate in this region, the electron particle flux shows
 qualitatively the same behavior in both cases.  

 In TE mode dominated plasmas -- as in our experimental scenarios --
 moving from pure ohmic to electron cyclotron heating leads to higher
 electron energy flux, since the collisional stabilization of the TE
 mode is less effective, and increasing the electron temperature
 gradient and electron-to-ion temperature ratio enhances the energy
 flux even further. On the other hand the turbulent electron particle
 flux can remain approximately unchanged as the TE growth rate
 decreases with increasing electron-to-ion temperature ratio,
 balancing the opposite effect of the lower collisionality and the
 higher electron temperature gradient.  It leads to the conclusion
 that the experimentally observed slight flattening of the electron
 density profile may have other reasons, e.g.  the strength and period
 of sawteeth in the central region can be different and this can have
 implications on the density profile.

 The simulations indicate that the impurity convective flux is
 negative in both the EC and OH cases, but it is significantly lower
 in the EC case. Furthermore the impurity diffusion coefficient is
 lower in that case.  As a consequence, the impurity peaking factor is
 lower in the EC case, however according to the simulations it does
 not change sign when electron cyclotron heating is applied.  A sign
 change in the peaking factor is therefore probably due to some
 additional physical mechanism, not accounted for in the linear
 gyrokinetic simulations. Recent work shows that impurity poloidal
 asymmetries may lead to a reduction or even sign change in the
 peaking factor \cite{poloidal}. Poloidal asymmetries may arise due to
 large pressure or temperature gradients if the plasma is sufficiently
 collisional \cite{fh2}, and in this case it could be caused of the
 large temperature gradient due to EC heating. Finally, impurity
 accumulation is affected also by neoclassical processes, and the
 neoclassical impurity inward pinch is expected to be reduced in the
 presence of ECRH due to the flatter ion density profile.


\section*{Acknowledgments}
The authors gratefully acknowledge helpful conversations with
V. Krupin and V. A. Vershkov, and would like to thank J. Candy for
providing the {\sc gyro} code. The authors acknowledge the work of the
T-10 experimentalists, who provided the information about plasma
parameters.  This work was funded by the European Communities under
Association Contract between EURATOM and {\em
  Vetenskapsr{\aa}det}. The views and opinions expressed herein do not
necessarily reflect those of the European Commission. One of the
authors, S. M., acknowledges support from the Wenner-Gren Foundation.


\end{document}